\documentstyle[psfig,prb,aps]{revtex}
\begin{document}

\headheight-1.0cm
\headsep0.0cm
\topmargin0.6cm
\textheight24cm

\draft

\twocolumn[\hsize\textwidth\columnwidth\hsize\csname
@twocolumnfalse\endcsname

\title{Coadsorption of CO and O on Ru\,(0001): \\
A structural analysis by density functional theory}
\author{C. Stampfl and M. Scheffler}
\address{
Fritz-Haber-Institut der Max-Planck-Gesellschaft,
Faradayweg 4-6, D-14195 Berlin-Dahlem, Germany\\
}
\date{\today}

\maketitle

\begin{abstract}
Knowledge of the atomic geometry of a surface is a 
prerequisite for any detailed understanding of 
the surface's electronic structure and chemical properties.
Previous studies have convincingly demonstrated that density functional
theory
(DFT) yields accurate surface atomic geometries
and that reliable predictions concerning stable and metastable phases
can be made on the basis of the calculated energetics.
In the present work we use DFT to investigate the
atomic structure of four ordered coadsorbate phases of
carbon monoxide and oxygen on Ru\,(0001). All of the structures
have a $(2 \times 2)$ periodicity with differing concentrations of
CO molecules and O atoms. For two of these phases
dynamical low-energy electron diffraction (LEED)
intensity analyses have been performed
and the  agreement between our DFT-
and the LEED-determined structures
is found to be very good.
We predict the atomic geometry of the third phase for which
no structural determination based on experiments has been made  to date.
We also
predict the stability of a new ordered mixed phase.
\end{abstract}

\vskip2pc]

\section{Introduction}
The coadsorption system CO,O/Ru\,(0001) represents
a well-studied model system~\cite{hoffmann,kostov}
for the investigation of the interaction
and behavior of coadsorbed species that can give insight
into physical
processes that are relevant to
heterogeneous catalysis~\cite{ertl,goodman,somorjai}.
Quite clearly, in order to describe and understand chemical reactions
at surfaces and the surface's electronic structure,
it is at first necessary to know the atomic geometry. 
It is only recently that
the detailed atomic structure of some of the phases of CO,O on Ru\,(0001)
have been determined,
and new ones discovered.
Depending on the experimental preparation, coadsorption of
CO and O on Ru\,(0001) have been reported to form the following phases:
$(2 \times 2)$-(1O+1CO)~\cite{kostov,narloch1},
$(2 \times 2)$-(2O+1CO)~\cite{narloch2},
and $(2 \times 2)$-(1O+2CO)~\cite{schiffer}.
In the rest of the paper we omit indicating the periodicity since
for all of the coadsorption structures considered, it is the same.
For the first two structures, 
low-energy electron diffraction
(LEED) intensity analyses have been performed: In the first phase,
the O atoms occupy hcp sites and the CO molecule adsorbs in the
on-top site; interestingly, a
tilt of the CO molecular axis of 12.6$^{\circ}$ was identified.
In the second phase, a restructuring induced by CO adsorption
of the O atoms of the $(2 \times 1)$~\cite{lindroos}
phase occurs: half of the O atoms,
initially occupying the hcp sites, switch to fcc sites and CO
adsorbs again in the on-top site; the other half of the O atoms remain
in hcp sites.
For the (1O+2CO) structure, to our knowledge, there has been no 
LEED intensity analysis, but infrared absorption spectroscopy
(IRAS) and X-ray photoelectron spectroscopy (XPS)
experiments~\cite{schiffer} indicate
that the O atoms occupy hcp sites and the CO molecules occupy
on-top and fcc sites. For this structure,  it was found that
in order to achieve the higher coverage of CO, notably higher exposures
were necessary, i.e. $\approx 10^{5}$~Langmuirs.
In correspondence, the sticking coefficient, which is
initially close to
one, drops by a factor of 60 at CO coverages greater than $\approx$0.25.
That is, initially there is a steep CO uptake curve,
thereafter there is a kink and it rises only slowly with exposure.

Previous studies of the structure and stability of adsorbate phases
using density-functional theory (DFT) have demonstrated that atomic
geometries and stable and metastable structures can be accurately
predicted (see, e.g.~\cite{alkali,stampfl,over,other1}). 
In the present work
we perform DFT calculations~\cite{stumpf}
for each of the identified phases in order to investigate,
at first, the atomic structure.
This work represents one of the first DFT studies of such
coadsorbate systems where quantitative comparison with structures
determined by LEED are made.
In a subsequent publication we will report on the energetics
and further details of the electronic structure.
We compare our calculated atomic geometries for the (1O+1CO)
and (2O+1CO) phases with the results of the
LEED intensity analyses, and predict that of the to date
undetermined (1O+2CO) phase.
In addition, we investigated a hypothetical
(3O+1CO) structure,
which we predict to represent a new high density stable phase.
These mixed O,CO/Ru\,(0001) surface structures are depicted in
the top panel of Fig.~1.

\section{Calculation method}
We use the generalized gradient approximation
(GGA) for the exchange-correlation functional~\cite{perdew}
and the supercell approach to model the surface structures, which
consists of four Ru layers. The O atoms and CO molecules
are adsorbed on one side of the slab
\cite{stumpf}.
We use {\em ab initio}, fully separable pseudopotentials~\cite{troullier}
where the GGA is employed for all atoms~\cite{martin}, and
relativistic effects taken into account for the Ru atoms (using
weighted spin-averaged pseudopotentials).
The positions of the C and O atoms, and the Ru atoms in the top
two layers are relaxed. The
energy cutoff is 40~Ry and there
are three special {\bf k}-points in the surface Brillouin
zone~\cite{cunningham}.

\section{O on Ru\,(0001)}
In an earlier publication~\cite{stampfl} we reported results for
ordered adlayers of O on Ru\,(0001); some of these results
will be used
as reference structures for comparison~\cite{note}.
In addition, we perform calculations for
an artificial ``honey-comb'' $(2 \times 2)$-2O structure
in which one O atom in the surface unit cell
occupies an hcp site and the other an fcc site.
The atomic geometry of the honey-comb structure is as for
the (2O+1CO) phase, but without the CO molecules.
The adsorption of oxygen on Ru\,(0001) under ultra high vacuum (UHV)
conditions
forms two ordered phases: $(2 \times 2)$-O~\cite{pfnuer} and
$(2 \times 1)$-O~\cite{lindroos} for coverages 1/4 and 1/2,
respectively. Recently two additional phases have been identified:
$(1 \times 1)$-O~\cite{over}
and $(2 \times 2)$-3O~\cite{kostov2,over2} structures for coverages
$\Theta$=1 and 3/4, respectively. Formation of the latter two phases
require introduction of higher concentrations of oxygen to the
surface either by employing very high gas pressures of O$_{2}$
or by using ``atomic oxygen'' via NO$_{2}$
dissociation. In all of the ordered phases
of  O on Ru\,(0001) that form in nature, O adsorbs in the hcp site.
This is the ``natural'' site, {\em i.e.} the site where also Ru 
would sit; it is
stabilized by a lowering of the occupied DOS, following the
trend that systems like to attain a chemically hard electronic structure,
{\em i.e.} a low density at the Fermi level.
Thus, essentially the same effect which stabilizes the
hcp structure for Ru over the fcc structure (see e.g.~\cite{skriver}).

\section{(1O+1CO)/Ru\,(0001)}
In the coadsorption system,
(1O+1CO), CO adsorbs in the on-top site
and O in an hcp site~\cite{narloch1}.
The structural parameters determined by LEED and DFT-GGA are given in
Tab.~I (compare Fig.~1).
The CO bond length determined by both methods is
$\approx$1.16~\AA\,, only
slightly longer than
that of the free molecule which is 1.15~\AA\, (as calculated in the
present work and also as obtained experimentally,
see e.g. Ref.~\cite{pople}) and
only slightly shorter than that on the clean surface
($\approx$1.17~\AA\,~\cite{herbert}).
The LEED-determined C-Ru distance is 1.93~$\pm$0.06~\AA\,  (just the same
as on the clean surface~\cite{herbert}) and the DFT-GGA value is
1.950~\AA\,.
The position of the Ru atom to which CO is adsorbed is displaced
slightly {\em inwards} by $\approx$0.06~\AA\, (by LEED and DFT-GGA)
 relative to the top-most Ru atoms in the unit
cell. As discussed in Ref.~\cite{narloch1}, this is in apparent contrast
to the behavior of CO  on the clean surface where,
the Ru atom is ``pulled'' outwards by 0.07~\AA\,.
We point out, however, that with respect to the $(2 \times 2)$-O structure,
the same Ru atom to which CO would adsorb is {\em already} displaced
inwards by 0.05~\AA\, as determined by LEED~\cite{lindroos}
(0.043~\AA\, by DFT-GGA~\cite{stampfl}),
relative to the other three surface Ru atoms.
Thus CO adsorption does not actually alter the relative position
of the Ru atom very much.
Compared to the vertical distance between the center of mass of the first
and second Ru layers of the $(2 \times 2)$-O structure~\cite{lindroos},
however, CO has induced a slight expansion, of 1.2~\% as determined
by LEED and 1.35~\% by DFT-GGA.

The LEED intensity analysis~\cite{narloch1}
 also identified a tilt of
12.6$^{\circ}$ of the CO molecular axis in a direction
between two adsorbed O atoms;
it was postulated that the tilt was {\em static},
rather than dynamic.
We investigated such a tilting of the CO axis
by initially taking the identified
tilt and allowing atomic relaxation.
The final tilt angle was found to be
6$^{\circ}$.
The difference to the LEED result lies mainly
in the lateral position of the C atom:
LEED finds a very small displacement (0.05$\pm$0.11~\AA\,)
while DFT-GGA obtains a larger
value (0.141~\AA\,) giving rise to a tilt of the C-O-Ru
{\em complex} rather than just
of the C-O axis. The results in any case agree to within the
experimental error bars.
In addition we considered tilting of the CO axis {\em towards} an adsorbed
atom.
We took the initial tilt to be that of 12.6$^{\circ}$ as determined
by LEED. On relaxation of the atomic positions, we found again
a tilt of $\approx$6$^{\circ}$. On inspection of the energetics
of both the tilt-geometries
we found that the total energy was practically identical to the upright
configuration. 
Thus the potential energy surface (PES) is rather flat which
suggests that the CO molecule will strongly vibrate.

The corresponding electron density of the valence states
for this phase is shown in Fig.~1.
The oxygen atoms appear as the red, almost spherical features.
Both adsorbate species can be seen to
induce a redistribution of the electron density of the top-layer Ru atoms, where
the C-Ru bond is clearly seen.

\section{(2O+1CO)/Ru\,(0001)}
The coadsorption system (2O+1CO) is formed by CO
adsorption onto the
$(2 \times 1)$-O phase. CO induces a
restructuring of the O atoms such that
half of them
move into the fcc sites, while the other half remain in hcp sites. By
doing so, CO maintains its favored on-top adsorption site.
The structural parameters of this phase as determined by LEED\cite{narloch2}
and DFT-GGA
are given in Tab.~II (compare Fig.~1). It can be seen that
there is good general agreement.
Both LEED and DFT-GGA find a slightly shorter CO bond length
(i.e., of 1.15~\AA\, and 1.155~\AA\,, respectively)
than for either CO on the clean surface ($\approx$1.17~\AA\,) or in the
(1O+1CO) phase ($\approx$1.16~\AA\,).
There is, however,  a slight difference between the LEED and DFT-GGA
results: The LEED-determined  C-Ru
bond length is {\em longer} in this phase than for CO on the clean
surface or in the (1O+1CO) phase (1.98~\AA\, compared
to 1.93~\AA\,) indicating a weaker C-metal bonding.
The DFT-GGA calculations
find a similar, but somewhat {\em shorter} C-Ru bond length (1.922~\AA\,)
than in the (1O+1CO) phase (1.950~\AA\,), indicating
a somewhat {\em stronger} bonding of CO to the metal
in this phase. In a forthcoming publication we will
investigate the energetics of all the
ordered mixed coadsorption phases.

The trends in the rumpling and lateral shifts of the top layer Ru atoms
are agree well between LEED and DFT-GGA.
Similarly to the (1O+1CO) phase discussed above, the Ru atom
to which CO is bonded is displaced {\em inwards} relative to the
other top-layer Ru atoms, by 0.09~\AA\, (LEED) and by 0.04~\AA\, (DFT-GGA).
For the underlying O structure of the (2O+1CO)  phase,
which corresponds to the artificial
``honey-comb'' geometry described above,
we point out that a similar, but {\em larger}
 difference
already exists between the vertical positions of the
top-layer Ru atoms; namely, the same Ru atom to which CO would bond
is 0.154~\AA\,  further in towards the surface than the other Ru atoms in
the surface unit cell.
Therefore, relative to the position of the
Ru atoms of the {\em honey-comb structure}, the Ru atom to which CO bonds is
in fact displaced {\em outwards} by 0.114~\AA\, due to CO adsorption.

While the LEED study did not report any relaxation of the second Ru layer,
the DFT-GGA calculations identify small relaxations and a rumpling:
The Ru atom below the hcp O atom is
displaced 0.038~\AA\,
further in towards the bulk than the other three Ru atoms. These latter
three Ru atoms exhibit lateral
displacements of 0.013~\AA\, radially in towards each other.
The top Ru-Ru vertical displacement, using the center of mass
of each layer, is 0.6~\% larger
than that of the honey-comb geometry, which is due to CO adsorption.

From the valence electron density (Fig.~1) it can be seen that
the bonding of the hcp O atom to the metal appears to be
stronger than the fcc O atom; this is what we may expect on the
basis of the magnitude of the energy difference of between
the hcp and fcc sites in the stable $(2 \times 1)$ structure
for the same coverage~\cite{stampfl}.

\section{(1O+2CO)/Ru\,(0001)}
For the (1O+2CO) phase no LEED analysis exists
so far.
The calculated atomic geometry
is depicted in Fig.~1.
The CO bond length for CO in the on-top site is 1.157~\AA\, and for
CO in the fcc site the bond length is a longer 1.187~\AA\,.
For the top site,
the C-Ru bond length is 1.968~\AA\,, similar to that of  the
(1O+1CO) phase
(and slightly longer than that in the (2O+1CO) phase, which
was 1.922~\AA\, as determined by DFT-GGA). For the fcc site
the C-Ru bond length is 2.238~\AA\,,
notably longer than for CO in the on-top
site;
this, however, is expected due to the fact that it is
{\em three-fold} coordinated and not one-fold.
The calculations also identify small lateral displacements of 0.018~\AA\,
of the three top-layer Ru atoms in the direction towards the fcc site.
Also, a rumpling of the top Ru layer of 0.129~\AA\,
occurs as indicated in Fig.~1, where again the Ru atom to which
CO bonds in the on-top site is further in towards the bulk.
In this case, with respect to the $(2 \times 2)$-O structure, the Ru
atom is
displaced inwards by 0.086~\AA\, (i.e., 0.129$-$0.043~\AA\,).
In the second Ru layer, the Ru atoms below the hcp O atoms are 0.021~\AA\,
in further towards the bulk than the other three Ru atoms,
and again these three Ru
atoms display a lateral radial displacement towards each other of
0.023~\AA\,.
We hope this predicted atomic geometry can be confirmed by a LEED intensity
analysis.

From the cross-section shown of the valence electron density,
the bonding of the CO molecule in
the fcc site can be seen to be
weaker than for that of the on-top site. In the fcc site, however,
CO forms three bonds with the metal surface so it is expected
that they be longer and weaker.

\section{(3O+1CO)/Ru\,(0001)}
We have also studied the atomic geometry of a (3O+1CO)
structure, although, to date, such a phase
has not been 
reported to exist experimentally. Our results predict that the
structure is energetically stable, although the
adsorption energy is low: Starting from a 
$(2 \times 2)$ array of O vacancies in the full oxygen monolayer
the CO adsorption energy is 0.85~eV.
As mentioned above,
recently it has been shown that such a
$(2 \times 2)$-3O/Ru\,(0001) ``vacancy'' or
``hole'' structure in fact represents a new stable phase of O on
Ru\,(0001)~\cite{kostov2,over2}.
The stability of this phase was also indicated in our study
of the thermodynamics of the O/Ru adsorption system~\cite{kreuzer}.
In order to investigate the energetics of CO adsorption
into the vacant hcp site
we calculated the total energy of CO at various distances above this site.
We found that there is an energy
barrier of $\approx$0.3~eV to adsorption.
This implies that rather high CO pressures would be
required in order to realize this structure experimentally.
Providing these activation barriers can be overcome,
we propose that
this structure
represents a new ordered $(2 \times 2)$ mixed phase of CO and O
on Ru\,(0001).
The atomic geometry is depicted in Fig.~1.
Here the  CO bond length is 1.181~\AA\,, somewhat
longer than for CO in the on-top site, but similar to that of
CO in the fcc site for the previous structure discussed, which was
1.187~\AA\,.
In the (3O+1CO) system,
there are only negligible lateral shifts of the O atoms
and also only negligible rumpling of the top
Ru layer (i.e., all $<$0.005~\AA\,).
In the second Ru layer, however, the rumpling is 0.043~\AA\,.
In this case the Ru atom directly under
CO is displaced {\em outwards} (or upwards)
relative to the other three Ru atoms.
There are again only negligible lateral displacements in the second Ru
layer.
Due to CO adsorption, using the center of gravity of the layers,
the vertical
Ru-Ru distance is expanded by $\approx$2~\% relative to the
$(2 \times 2)$-3O O-vacancy structure.
Thus again, CO adsorption has induced an overall expansion of the
first Ru-Ru interlayer spacing.

The valence electron density in Fig.~1 shows that there
is the most disturbance of the Ru states for this structure.
With respect to CO adsorption, we see four clear maxima of the electron
density on the Ru atom below, while for the Ru atom that only forms
bonds with O atoms, two strong maxima appear which are polarized
away from the O-Ru bond.
This redistribution of Ru states is seen to give
rise to a build-up of electron density on top of the Ru atoms in the
top layer.

\section{Conclusion}
We have investigated mixed O, CO adlayers on Ru\,(0001).
We find very good overall agreement of the atomic geometry
of the (1O+1CO) and (2O+1CO) phases
with those determined by LEED intensity analyses. We predict the
to date undetermined structural geometry of the (1O+2CO) phase and
propose the existence of a new
high density phase, namely, (3O+1CO).
With respect to the latter structure,
adsorption of CO onto the $(2 \times 2)$-3O surface in order
to achieve this phase is found to be
activated. This, together with the notably reduced area for CO adsorption
on this surface,
suggests that high CO pressures would 
be required to create this phase experimentally.

\begin{onecolumn}

\begin{figure}
\vspace{-10mm}
\psfig{figure=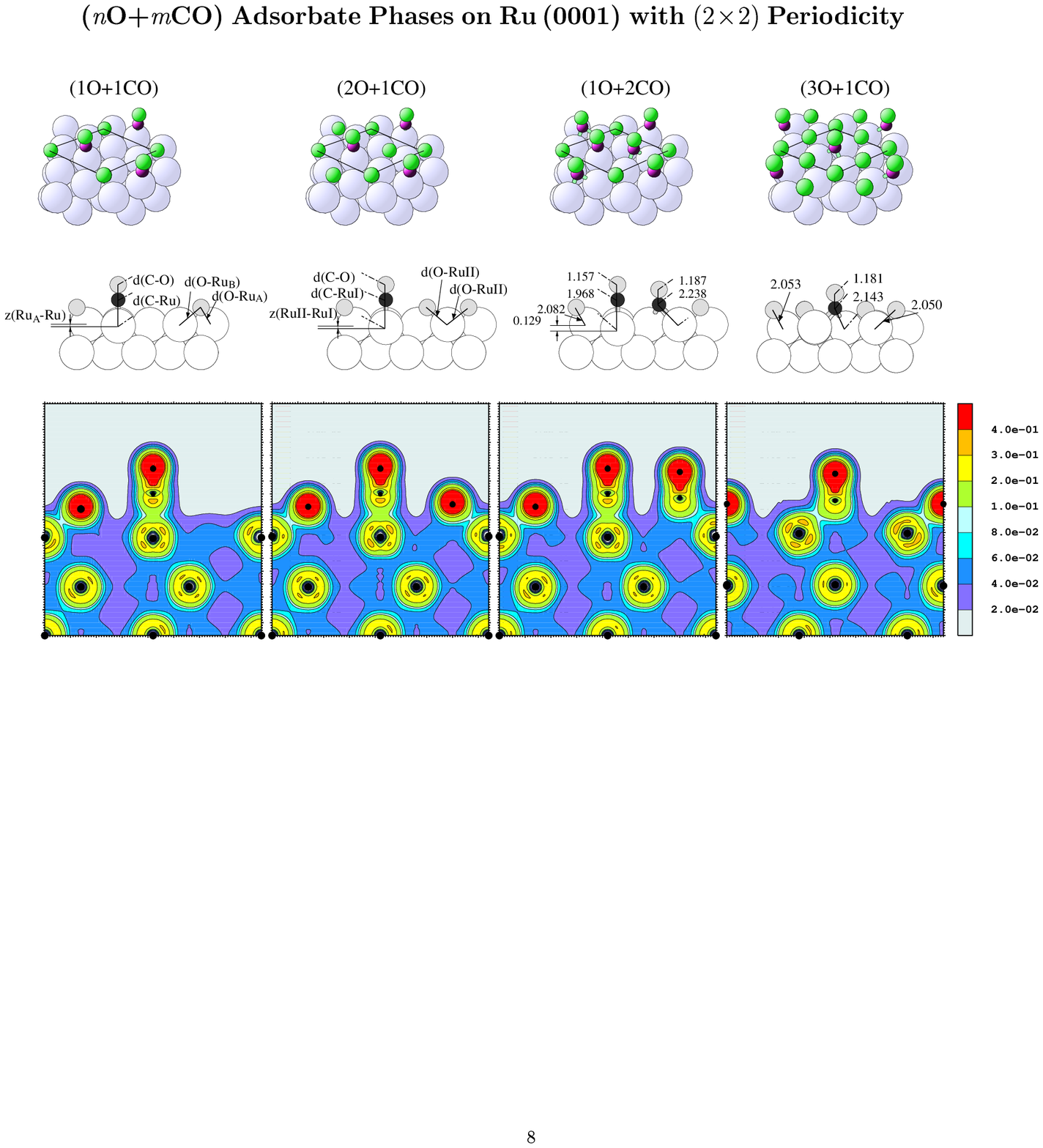,rheight=130mm}
\caption{ Perspective and  side views of the
various  phases of O and CO on Ru\,(0001).
Large and small (green and red) circles represent Ru, O, and C atoms,
respectively.  The lower panel shows the electron density
of the valence states.
The contour lines are given in $e$ bohr$^{-3}$.}
\end{figure}

\begin{table}
\begin{tabular}{c|c|c|c|c|c|c|c}
        & $d$(C-O) & $d$(C-Ru) & $d$(O-Ru$_{\rm A}$) &
$d$(O-Ru$_{\rm B}$) & $z$(Ru$_{\rm A}$-Ru) &
$\Delta_{\parallel}$(Ru$_{\rm A}$) & $\Delta_{\parallel}$(Ru$_{\rm B}$) \\
  & (\AA) & (\AA) & (\AA) & (\AA) & (\AA) & (\AA) & (\AA) \\
\hline
DFT-GGA &1.161  &1.950  & 2.109 & 2.166 & 0.057 &0.050  & 0.068 \\
LEED & 1.16$\pm$0.06 & 1.93$\pm$0.06 & 2.06$\pm$0.08 & 2.09$\pm$0.14 &
0.06$\pm$0.05 & 0.16$\pm$0.10 & 0.10$\pm$0.10 \\
\end{tabular}
\caption{Structural parameters obtained by DFT-GGA and by
LEED~\protect\cite{narloch1} for
the (1O+1CO) phase (see Fig.~1).}
\end{table}

\begin{table}
\begin{tabular}{c|c|c|c|c|c|c|c}
        & $d$(C-O) & $d$(C-Ru) & $z$(O(fcc)-O(hcp)) &
$d$(O(hcp)-RuII) & $d$(O(fcc)-RuII) & $z$(RuII-RuI) &
$\Delta_{\parallel}$(RuII) \\
  & (\AA) & (\AA) & (\AA) & (\AA) & (\AA) & (\AA) & (\AA) \\
\hline
DFT-GGA & 1.155 & 1.922 & 0.157 & 2.080 & 2.122 & 0.040 & 0.039 \\
LEED & 1.15$\pm$0.04 & 1.98$\pm$0.08 & 0.19$\pm$0.09 & 2.05$\pm$0.16 &
2.05$\pm$0.16 & 0.09$\pm$0.02 & 0.08$\pm$0.08 \\
\end{tabular}
\caption{Structural parameters obtained by DFT-GGA and by
LEED~\protect\cite{narloch2} for
the (2O+1CO) phase (see Fig.~1).}
\end{table}

\end{onecolumn}

\begin{twocolumn}

\end{twocolumn}


\begin{references}
\bibitem{note}Exactly the same basis set was used for the calculation of
the ordered adlayers of O on Ru\,(0001).
\end{references}
\end{document}